# *In situ* high-pressure synchrotron x-ray diffraction study of CeVO$_4$ and TbVO$_4$ up to 50 GPa

D. Errandonea[1,*], R. S. Kumar[2], S. N. Achary[3], and A. K. Tyagi[3]

[1]*MALTA Consolider Team, Departamento de Física Aplicada-ICMUV, Universitad de Valencia, Edificio de Investigación, c/Dr. Moliner 50, Burjassot, 46100 Valencia, Spain*

[2]*High Pressure Science and Engineering Center, Department of Physics and Astronomy, University of Nevada Las Vegas,4505 Maryland Parkway, Las Vegas, Nevada 89154-4002, USA*

[3]*Chemistry Division, Bhabha Atomic Research Centre, Trombay, Mumbai 400085, India*

**Abstract:** Room temperature angle-dispersive x-ray diffraction measurements on zircon-type TbVO$_4$ and CeVO$_4$ were performed in a diamond-anvil cell up to 50 GPa using neon as pressure-transmitting medium. In TbVO$_4$ we found at 6.4 GPa evidence of a non-reversible pressure-induced structural phase transition from zircon to a scheelite-type structure. A second transition to an M-fergusonite-type structure was found at 33.9 GPa, which is reversible. Zircon-type CeVO$_4$ exhibits two pressure-induced transitions. First an irreversible transition to a monazite-type structure at 5.6 GPa and second at 14.7 GPa a reversible transition to an orthorhombic structure. No additional phase transitions or evidences of chemical decomposition are found in the experiments. The equations of state and axial compressibility for the different phases are also determined. Finally, the sequence of structural transitions and the compressibilities are discussed in comparison with other orhtovanadates and the influence of non-hydrostaticity commented.

---

[*] Author to whom correspondence should be addressed. Electronic mail: daniel.errandonea@uv.es.

I.  **Introduction**

Orthovanadates formed by lanthanides are of particular interest since a relatively high concentration of various lanthanides is present in the waste resulting from reprocessing of light water reactor spent fuel in the recent years. The main potential applications include host matrices for immobilization of certain types of radioactive wastes. The rare-earth orthovanadates AVO$_4$ (where A is a lanthanide) crystallize in the tetragonal zircon-type structure (space group: $I4_1/amd$, Z = 4) at ambient conditions [1], with exemption of dimorphic LaVO$_4$, which has a zircon-type structure or a monoclinic monazite-type structure (space group: $P2_1/n$, Z = 4) depending on the mode of preparation [2]. In the zircon structure, the vanadium atom is tetrahedrally coordinated, while the trivalent lanthanide is coordinated by eight oxygen atoms, forming a bidisphenoid. During the last decade, intensive investigations have been carried out on the structural evolution of the zircon-type vanadates under extreme conditions. In particular, x-ray diffraction [3 - 6], optical [7, 8], and Raman scattering measurements [4, 6, 8, 9, 10], and theoretical calculations [5, 6, 10, 11] have been carried out to understand the structural modifications induced by pressure. Orthovanadates with small A cations have been observed to undergo an irreversible zircon-to-scheelite (space group: $I4_1/a$, Z = 4) phase transition [3]. In some compounds like LuVO$_4$ and EuVO$_4$ a second transition induced by pressure has been detected. The second transition is from the tetragonal scheelite-type structure to a monoclinic fergusonite-like structure (space group: $I2/a$, Z = 4). However, in CeVO$_4$, a compound with a large rare-earth cation, a different structural sequence has been observed. According with Raman experiments, under compression CeVO$_4$ transforms from zircon to monazite under quasi-hydrostatic conditions [4]. In contrast under non-hydrostatic conditions the zircon-to-scheelite transition has been detected. Both, non-hydrostatic effects and the change of the trivalent cation have been shown to play a role also in the structural behavior of zircon-type arsenates, chromates, and phosphates [12, 13]. Both factors deserve to be further studied into detail. Here, in order to shed more light on the understanding of the mechanical and structural properties of zircon-type oxides, we report *in situ* high-pressure (HP) room-temperature (RT) x-ray diffraction studies of TbVO$_4$ and CeVO$_4$ up to a pressure of 50 GPa using neon (Ne) to create quasi-hydrostatic conditions. The present work contributes to deeper the understanding of pressure effects on the crystal structure of zircon-type oxides of both technological and geophysical importance. Among other things, geophysical relevance of zircon-type oxides is related to the fact that they are important accessory minerals in granitoids and rhyolites, and because of their presence in igneous rocks.

## II. Experimental details

The TbVO$_4$ and CeVO$_4$ samples used in the experiments were prepared by solid-state reaction of appropriate amounts of pre-dried Tb$_2$O$_3$ or CeO$_2$ (Indian Rare Earth Ltd. 99%) and V$_2$O$_5$ (Alfa-Aesar 99%). Homogeneous mixtures of the reactants were pelletized and heated at 800 °C for 24 h and then cooled to ambient temperature. Further, the pellets were reground and heated again at 1100 °C for 24 h. The samples obtained were characterized by powder x-ray diffraction data recorded on a Panalytical X-pertPro diffractometer using Cu $K_\alpha$ radiation. Single phase samples of zircon-type structure were confirmed in all. The refined unit-cell parameters for these phases are $a = 7.377(7)$ Å and c = 6.485(6) Å for CeVO$_4$ and $a = 7.166(7)$ Å and $c = 6.317(6)$ Å for TbVO$_4$. These values are in agreement with earlier reported values [6, 14].

In order to perform high-pressure studies, pre-pressed pellets of TbVO$_4$ or CeVO$_4$ were prepared using the finely ground powder obtained from the synthesized samples. Synchrotron powder x-ray diffraction experiments were performed up to 50 GPa. The pellets were loaded in a 100-μm hole of a rhenium gasket pre-indented to 30 μm in a symmetric diamond-anvil cell (DAC) with diamond-culet sizes of 300 μm. A few ruby grains were loaded with the sample for pressure determination [15] and Ne was used as the pressure-transmitting medium [16, 17]. At pressures higher than 4 GPa (solidification of Ne) the equation of state (EOS) of Ne was used to double check the pressure when the Ne peaks were detected [18]. Pressure differences between both methods were always smaller than 0.2 GPa. Angle-dispersive x-ray diffraction (ADXRD) experiments were carried out at Sector 16-IDB of the HPCAT, at the Advanced Photon Source (APS), with an incident wavelength of 0.4246 Å. The monochromatic x-ray beam was focused down to $10 \times 10$ μm$^2$ using Kickpatrick-Baez mirrors. The images were collected using a MAR345 image plate located 350 mm away from the sample and then integrated and corrected for distortions using FIT2D [19]. The structure solution and refinements were performed using the POWDERCELL [20] and GSAS [21] program packages.

### III. Results and Discussion

#### A. TbVO$_4$

Figure 1 shows a selection of diffraction patterns of TbVO$_4$ measured at different pressures up to 50 GPa. Assuming a zircon-type structure, all the patterns observed between ambient pressure and 4.4 GPa can be well indexed. Earlier a structural transition around 33 K has been reported in temperature dependent studies on TbVO$_4$ [14]. This structural transition has been assigned to the cooperative Jahn-Teller distortion of Tb$^{3+}$ leading to the distorted TbO$_8$ bidisphenoid with a minor reduction (0.7 %) of unit-cell volume. However, in this pressure study the Jahn-Teller transition is not observed. At higher pressure, viz at 6.4 GPa we observed the appearance of extra peaks in addition to the peaks associated to the zircon structure. The intensities of the emerging peaks gradually increase from 6.4 to 8.8 GPa. At this pressure all the peaks of the low-pressure phase are not detectable anymore which indicate completion of structural transition. The onset of the transition is at 6.4 GPa, and the transformation is fully completed up to 8.8 GPa. Within this range of pressure, the low- and high-pressure phases coexist. The transition pressure agrees with that obtained from Raman experiments [8]. The transition is gradual but the coexistence range is reduced in comparison with that observed on $EuVO_4$ and $LuVO_4$ [3]. We think the coexistence range is reduced in the present experiment due to the generation of nearly quasi-hydrostatic conditions through the use of Ne as pressure medium. In contrast, in the $EuVO_4$ and $LuVO_4$, experiments silicone oil was the pressure medium, generating non negligible uniaxial stresses, which could modify the onset of the transition and the pressure range of phase coexistence [22, 23].

From 8.8 to 30.3 GPa there is no more changes in the diffraction patterns of $TbVO_4$, but at 33.9 GPa we found a broadening of the peaks and the appearance of additional weak peaks. These facts indicate that a second pressure-driven phase transition takes place. Upon further compression no additional changes are observed in the diffraction patterns. Therefore $TbVO_4$ undergoes two transitions up to 50 GPa. In addition, we did not find any evidence of pressure-induced chemical decomposition. Regarding the second phase transformation, we found it is reversible with a large hysteresis. When releasing pressure from 49.2 GPa the transition reversion occurs at 26 GPa. On the contrary, the other transition appears to be non-reversible as illustrated by the diffraction pattern measured at 0.3 GPa on pressure release. This pattern is quite similar to the patterns measured under pressure increase once the first transition occurs instead of resembling a pattern from zircon phase. This fact is in good agreement with the irreversible changes detected in Raman experiments [8]. In this work, the high-pressure phase was assigned to a scheelite-type structure by means of x-ray diffraction studies performed on quenched samples. In our case, after a deep examination of the diffraction patterns measured from 8.8 to 30.3 GPa, we found that they can be well indexed considering a scheelite-type structure. From a Rietveld refinement, at 8.8 GPa we obtained the unit-cell parameters $a = 4.980(5)$ Å and $c = 11.086(9)$ Å for the high pressure phase. The good quality of the refinements is illustrated by the fact that residuals are so small that it is difficult to differentiate in Fig. 1 between the measured and calculated diffraction patterns. In the scheelite structure Tb and V are located at *4b* and *4a* Wyckoff positions which are fixed. The O atoms are at the position *16f* with coordinates (0.2702, 0.1775, 0.0843). The transition from zircon to scheelite involves a volume collapse of approximately 12% and consequently is of first order. An identical transition was previously reported in many zircon-type oxides and in particular in rare-earth vanadates with small size lanthanides [3]. Upon pressure release we obtained for the scheelite-type phase at 0.3 GPa the following unit-cell parameters: $a = 5.054(5)$ Å and $c = 11.276(9)$. These values compare very well with the values reported at ambient pressure by Duclos *et al.* [8]; $a = 5.09$ Å and $c = 11.3$ Å.

Regarding the changes observed in the diffraction patterns beyond 33.9 GPa, they are consistent with the occurrence of a scheelite-to-fergusonite structural transition as previously found for $LuVO_4$ and $EuVO_4$ [3]. However, in these two compounds the scheelite to fergusonite transition was detected at a much lower pressure (22 – 25 GPa). Again the differences in transition pressures could be probably caused by the use of different pressure medium (non-hydrostaticity trigger the transition at lower pressure in $LuVO_4$ and $EuVO_4$ and probably in all orthovanadates). Unfortunately the broadening of diffraction patterns beyond 33 GPa does not allow the performance of a structural refinement. However, a LeBail analysis allowed us to correctly index the diffraction patterns collected beyond 33 GPa considering a *M*-fergusonite-type structure. Considering this structure, the following unit-cell parameters $a$ = 4.88(1) Å, $b$ = 10.70(2) Å, $c$ = 4.70(1) Å, and β = 91.0(2)° are obtained for this phase at 40.6 GPa. According to this result, no noticeable volume change occurs at the scheelite to fergusonite transition. To close this part of the discussion we would like to add that the zircon-schelite-fergusonite sequence found in TbVO$_4$ does not show any evidence of induction of cooperative Jahn-Teller effects as observed in the phase transitions driven at low temperature [14].

From our experiments we extracted the pressure evolution of the unit-cell parameters for zircon- and scheelite-type phases. The results are summarized in Fig. 2. For comparison we include the results obtained for the zircon phase from energy-dispersive x-ray diffraction (EDXRD) studies performed using a large-volume press (LVP) [24]. Both data compare well in the pressure range where comparable with the only difference that the non-hydrostatic conditions of the LVP experiment leads to a small reduction of compressibility as previously found in related ternary oxides [25, 26]. As can be seen in the figure, the compression of the zircon-type structure is non-isotropic, the $c$ axis being the less compressible axes. This fact can be better seen in Fig. 3 where we show the evolution of axial ratios for different phases. As a consequence of the non isotropic compression, the axial ratio (*c/a*) gradually increases from 0.881 at ambient pressure to 0.891 at 6.4 GPa. Regarding the unit-cell parameters of the scheelite-type structure, we found that compression is also anisotropic, the $c$ axis being the most compressible axes as in most scheelites [27]. In particular, the axial ratio *c/a* decreases non-linearly from 2.231 at ambient pressure to 2.185 at 30.3 GPa. It is important to note here that pressure effects on axial ratios can be affected by the presence of uniaxial stresses due to shear deformation. Therefore, since our experiments were performed using Ne as pressure (and consequently uniaxial stresses are minimized), the pressure evolution of axial ratios here reported is more accurately than that obtained from previous studies [24].

From the pressure dependence of the lattice parameters, the unit-cell volumes of zircon and scheelite phases of TbVO$_4$ as a function of pressure were calculated. The results are summarized in Fig. 4. We have analyzed the volume changes using a third-order Birch-Murnaghan EOS [28]. The obtained EOS parameters for zircon phase are: $V_0$ = 324.4(9) Å$^3$, $B_0$ = 122(5), and $B_0´$ = 6.2(5), these parameters being the zero-pressure volume, bulk modulus, and its pressure derivative, respectively. The bulk modulus of zircon-type TbVO$_4$ is comparable with the value obtained from the evaluation of elastic constants – 126 GPa [8] - and from EDXRD experiments - 121(5 GPa) [24]. However, EDXRD experiments underestimate considerably $B_0´$ - 4.1 GPa. This difference is caused by the underestimation of compression associated to non-hydrostatic conditions generated in the experiments by the LVP setup employed. The EOS parameters for scheelite phase are: $V_0$ = 288.2(6) Å$^3$, $B_0$ = 163(9) GPa, and $B_0´$ = 5.8(6). The EOS fits for both phases are shown as solid lines in Fig. 4. As in many other vanadates [3], the bulk modulus of the scheelite phase is more than 30% larger than that of the zircon phase. Empirical models have been developed for predicting the bulk moduli of zircon-structured and scheelite-structured ABO$_4$ compounds [29]. In particular, the bulk modulus of TbVO$_4$ can be estimated from the charge density of the TbO$_8$ polyhedra using the relation $B_0 = 610\, Z_i/d^3$, where $Z_i$ is the cationic formal charge of terbium, $d$ is the mean Tb-O distance at ambient pressure (in Å), and $B_0$ is given in GPa. Applying this relation a bulk modulus of 134(25) GPa is estimated for the low-pressure phase of TbVO$_4$ and a bulk modulus of 161(29) GPa is estimated for the scheelite-type phase. These estimations reasonably agree with the values obtained from our experiments and confirm that the scheelite-type phase is less compressible than the zircon-type phase.

**B. CeVO$_4$**

Figure 5 shows a selection of diffraction patterns of CeVO$_4$ measured at different pressures up to 50 GPa. The behavior is qualitatively different to that of TbVO$_4$ and other vanadates with small-size lanthanide cations. From ambient pressure up to 2.9 GPa all diffraction patterns can be well assigned to the zircon structure. At 3.9 GPa we detected the presence on many extra peaks, which are indicative of the onset of a phase transition. At 5.6 GPa the transition is fully completed. The pattern corresponding to the HP phase is totally different than those of the HP phases of TbVO$_4$; i.e. the HP phase of CeVO$_4$ should have a different crystal structure. At 5.6 GPa and higher pressures some Bragg reflections associated to the pressure medium (Ne) are observed. However, they are easy to identify and do not preclude the identification of the HP phases of CeVO$_4$. From 5.6 to 12.3 GPa there are no additional changes in the diffraction patterns; only the reflections move to higher angles due to the unit-cell reduction caused by compression. At 14.7 GPa important changes takes place in the patterns indicating that a second phase transition occurs. Upon further transition we cannot identify additional structural changes up to 45 GPa. As in TbVO$_4$, chemical decomposition is not induced by pressure. Upon decompression, the second transition is reverted but the transition occurring at 3.9 GPa is irreversible.

In previous studies it has been reported that both the scheelite and monazite structures can be achieved in CeVO$_4$ upon compression [6, 30]. In our case, in an experiment preformed under nearly hydrostatic conditions, the diffraction patterns of the HP phase can be undoubtedly assigned to the monoclinic monazite-type structure. A Rietveld refinement of the pattern collected at 5.6 GPa provide the following unit-cell parameters for monazite structure: $a = 6.877(6)$ Å, $b = 7.116(7)$ Å, $c = 6.624(6)$ Å, and $\beta = 104.5(3)°$. In this structure all the atoms are located at Wyckoff positions *4e*, being their coordinates: Ce (0.2787, 0.1580, 0.1017), V (0.3026, 0.1647, 0.6137), $O_1$ (0.2453, 0.0070, 0.4440), $O_2$ (0.3810, 0.3308, 0.4968), $O_3$ (0.4750, 0.1065, 0.8090), and $O_4$ (0.1261, 0.2166, 0.7183). At ambient pressure, in a quenched sample we obtained the following parameters for monazite: $a = 7.001(7)$ Å, $b = 7.221(7)$ Å, $c = 6.703(6)$ Å, and $\beta = 105.7(3)°$. These structural parameters agree with those previously reported by Panchal *et al.* [6] and imply a volume collapse of 9.6% at the transition. Both the volume contraction and the irreversibility of the transitions indicate it is of first order. An important fact to remark for monazite is that it has a Ce atom coordinated by nine oxygen atoms. Therefore, the phase transition involves a coordination increase of Ce from 8 to 9. The phase transformation is accomplished by breaking a Ce-O bond in zircon and adding two Ce-O bonds into monazite.

We have made also efforts to identify the crystal structure of the second HP phase. After considering several candidate structures, we found that the patterns measured beyond 14.7 GPa can be assigned to an orthorhombic structure with space-group symmetry $P2_12_12_1$. In particular, a LeBail analysis allows indexing the pattern measured at 14.7 GPa with a $CaSeO_4$-type structure ($a = 6.88$ Å, $b = 5.88$ Å, and $c = 6.76$ Å space group: $P2_12_12_1$, $Z = 4$) similar to the high-pressure structure recently found in $BaSO_4$ [31]. The orthorhombic structure is a distortion of barite and its appearance as post-monazite phase in $CeVO_4$ is consistent with the fact that barite-related structures have been found as HP phases in different monazites [13]. Barite and post-barite structures have been reported for high-pressure phases of $CaSO_4$ [32], $BaSO_4$ [31]. In case of anhydrite the barite-related structure has been reported as orthorhombic (Pbnm) phase through a monazite-type phase at high-pressure and high-temperature conditions, while a distorted-barite related phase ($P2_12_12_1$) is formed at high pressure itself in $BaSO_4$ [31]. A related structure is also found as a post-barite phase in $AgClO_4$ [33]. According with the orthorhombic HP structure we found, the volume change at the second transition is of approximately 14%. In addition the orthorhombic structure will involve an increase of the Ce coordination to 12, while the V coordination remains equal to 4. Thus, the structural sequence implies a gradual increase of the Ce coordination from 8 to 9 and to 12. Besides, the structural change is also reflected in the packing of the coordination polyhedral units, viz normal edge shared $CeO_8$ units of the zircon-type lattice transforms to the edge and face shared $CeO_{12}$ units of orthorhombic barite-related structure. Also close packing is resulted from the sharing of six $CeO_{12}$ units to one $CeO_{12}$ in contrast to the four of $CeO_8$ units located around $CeO_8$ of zircon structure. In contrast in $TbVO_4$, in the three structures observed the Tb and V coordination are not modified, being 8 and 4, respectively.

To understand the distinctive behavior of $CeVO_4$ under compression we will benefit of crystal-chemistry arguments [27]. In particular, it will be helpful to remember that $TbPO_4$ also undergoes the zircon to monazite transition [34]. Rare-earth phosphates with small-lanthanide cation crystallize in the zircon structure, but those whose lanthanide cation has an ionic radius larger than Tb crystallize in the monazite structure characteristic of $CePO_4$. Upon compression phosphates with smaller lanthanides, like $LuPO_4$ and $YbPO_4$ [35] transform from zircon to scheelite, but those closer to the zircon-monazite stability border accommodate stresses transforming from zircon to monazite, e.g. $TbPO_4$ and $ErPO_4$ [13, 35], which is much more flexible structure than scheelite [36]. An analogous behavior is expected for the vanadates with the only difference that at ambient conditions the frontier between zircon and monazite is located between Ce and La. Therefore it is reasonable to expect that most rare-earth vanadates would transform from zircon to scheelite as indeed do all compounds going from $LuVO_4$ to $EuVO_4$ [3]. On the other hand a compound with a large lanthanide like $CeVO_4$ would transform from zircon to monazite taking the ambient pressure structure of $LaPO_4$. According with recent optical and theoretical studies apparently this is the case also of $NdVO_4$ [7]. This suggests that $PrVO_4$, with the radius of Pr being between those of Ce and Nd, is also expected to undergo a transition to monazite upon compression. Then, among vanadates, only the behavior of $SmVO_4$ remains open. $SmVO_4$ is known to undergo a transition at 5.8 GPa [37], but still needs to be clarified whether it will follow the behavior of its small-lanthanide partners or its large-lanthanide partners.

From our experiments we also extracted the pressure dependence of the unit-cell parameters for zircon- and monazite-type phases. The results are summarized in Fig. 6. For comparison we include the results obtained using a 4:1 methanol-ethanol mixture as pressure medium. Both data compare well in the pressure of stability of the zircon phase. For the monazite phase differences smaller than 5% are observed. They are probably caused by the impoverishment of the quasi-hydrostatic conditions of the experiments in the experiment performed under 4:1 methanol ethanol. The presence of non-hydrostatic stresses lead to a peak broadening making difficult the accurate determination of unit-cell parameters. As can be seen in the figure, the compression of the zircon-type structure is non-isotropic, the *c* axis being again the less compressible axes. As a consequence of this, the axial ratio *c/a* gradually increases from 0.878 at ambient pressure to 0.887 at 4 GPa. Regarding the unit-cell parameters of the monazite-type structure, we found that compression is also anisotropic, the *a* axis being the most compressible axes. The evolution with pressure of the axial ratios for both phases can be seen in Fig. 3. The anisotropic compression of the monazite phase fully agrees with that of monazite-type LaPO$_4$ and CePO$_4$ [13, 38]. On the other hand, as observed in LaPO$_4$ [13], we also found that the β angle considerably decreases upon compression. This tendency to symmetry enhancement is consistent with the transformation to an orthorhombic structure at higher pressures.

From the pressure dependence of the lattice parameters, the unit-cell volumes of zircon and monazite phases of CeVO$_4$ as a function of pressure were calculated. The results are summarized in Fig. 7. We have analyzed the volume changes using a third-order Birch-Murnaghan EOS [28]. For the zircon phase there not enough data points to fit the three parameters of the EOS, therefore we fixed $V_0 = 352.9$ Å$^3$ and $B_0' = 4$ to perform the fit obtaining $B_0 = 125(9)$ GPa. This value is 5% larger than the value of 119 GPa [8] obtained from previous experiments [6]. The reduction of compressibility here observed is consistent with the use of noble gas a pressure-transmitting medium as previously found in related compounds [25, 26] and here for TbVO$_4$. However, it should be consider, that differences are comparable with error bars. In addition, the limited data points collected for the zircon phase in both works could influence the bulk moduli determination. The EOS parameters for monazite phase are: $V_0 = 326.2(8)$ Å$^3$, $B_0 = 133(5)$ GPa, and $B_0' = 4.4(6)$. The value of $B_0$ is 6% smaller than the value obtained from Ref. [6]. The present value is much more precise because given the lower quality of diffraction patterns in Ref. [6] it was assumed that the β angle was constant, a hypothesis we proved to be not appropriate. The EOS fits for both phases are shown as solid lines in Fig. 7. Note that in contrast with 30% increase of bulk modulus at the zircon-scheelite transition of TbVO$_4$, in the zircon-monazite transition the increase of the bulk modulus is only 6%. As described when discussing TbVO$_4$, empirical models have been developed for predicting the bulk modulus of zircon-structured oxides [29]. In particular, the bulk modulus of CeVO$_4$ can be estimated from the charge density of the CeO$_8$ polyhedra and the cationic formal charge of cerium, being the estimated value 120(25) GPa. Again, this empirical model works well for zircon structure. The reason for it is that contraction occurs basically because of the reduction of lanthanide-oxygen bonds. This model works well for zircon and scheelite, but it has been proven to be not correct for the monazite structure where also the V-O bonds contribute to volume reduction [13].

## IV. Conclusions

We have carried out room-temperature ADXRD experiments in the TbVO$_4$ and CeVO$_4$ using Ne as pressure medium up to 50 GPa. We detected the occurrence of two phase transitions in each compound. On the other hand, we did not find any evidence of chemical decomposition or induction of cooperative Jahn-Teller effects up to 50 GPa. The structural sequence of TbVO$_4$ is zircon-scheelite-fergusonite, as in most vanadates with small-size lanthanide cations, being the transition pressures 6.4 and 33.9 GPa. The structural sequence of CeVO$_4$ is zircon-monazite-CaSeO$_4$-type, being the transition pressures 5.6 and 14.7 GPa. The scheelite phase of TbVO$_4$ and the monazite phase of CeVO$_4$ are recovered at ambient pressure after decompression. Reasons for the different structural sequences are provided based upon crystal-chemistry arguments. The EOS and axial compressibility of each compound is also determined. In particular, we found that CeVO$_4$ is the most compressible zircon-type vanadate and that in the low- and high-pressure phases of both compounds compression is non isotropic. Finally our results are compared with previous studies to show that non-hydrostatic conditions could affect the high-pressure behavior of vanadates.

**Acknowledgments**

The authors thank for the financial support from Spanish MICCIN (Grants MAT2010-21270-C04-01 and CSD2007-00045). This work was performed at HPCAT (Sector 16), Advanced Photon Source (APS), Argonne National Laboratory. HPCAT is supported by CIW, CDAC, UNLV and LLNL through funding from DOE-NNSA, DOE-BES and NSF. APS is supported by DOE-BES, under DEAC02-06CH11357. The UNLV HPSEC was supported by the U.S. DOE, National Nuclear Security Administration, under DE-FC52-06NA26274.

**Figure Captions**

**Figure 1:** Selection of x-ray diffraction patterns measured in TbVO$_4$ at different pressures. Pressures are indicated in the figure as well as the structure assigned to each pattern. (r) denotes those patterns collected on pressure release. Vertical ticks indicate the calculated position for Bragg reflections. In the pattern collected at 8.8 GPa for the HP phase we show the experimental data (dots), the refined pattern (solid line), and the residual of the refinement (dotted line).

**Figure 2:** Pressure evolution of the unit-cell parameters of the zircon-type and scheelite-type phases of TbVO$_4$. To facilitate the comparison for the high-pressure phase we plotted *c*/2 instead of *c*. Lines: quadratic fits. Symbols: experiments. Solid squares: zircon, solid circles: scheelite, empty circles: scheelite upon decompression, and solid diamonds: zircon data from Ref. [24]. The vertical dashed line indicates the transition pressure.

**Figure 3:** Axial ratios for different phases of TbVO$_4$ and CeVO$_4$. Symbols: experiments from this work. Lines: linear fits. Solid symbols: compression data. Empty symbols: decompression data.

**Figure 4:** Pressure-volume relation in TbVO$_4$. Symbols: experiments. Lines: EOS fits. Solid squares: zircon, solid circles: scheelite, empty circles: scheelite upon decompression, and solid diamonds: zircon data from Ref. [24]. The vertical dashed line indicates the transition pressure.

**Figure 5:** Selection of x-ray diffraction patterns measured in CeVO$_4$ at different pressures. Pressures are indicated in the figure as well as the structure assigned to each pattern. (r) denotes those patterns collected on pressure release. Vertical ticks indicate the calculated position for Bragg reflections. In the pattern collected at 5.6 GPa for the HP phase we show the experimental data (dots), the refined pattern (solid line), and the residual of the refinement (dotted line).

**Figure 6:** Pressure evolution of the unit-cell parameters of the zircon-type and monazite-type phases of $CeVO_4$. Lines: quadratic fits. Symbols: experiments. Solid squares: zircon, solid circles: monazite, solid diamonds: zircon data from Ref. [6], and solid triangles: monazite data from Ref. [6]. Empty circles represent data obtained in this work upon decompression. The vertical dashed line indicates the transition pressure. The inset of the figure shows the evolution of the β angle of monazite.

**Figure 7:** Pressure-volume relation in $CeVO_4$. Symbols: experiments. Lines: EOS fits. Solid squares: zircon, solid circles: scheelite, solid diamonds: zircon data from Ref. [6] and solid triangle: monazite data from Ref. [6]. Empty circles represent data obtained in this work upon decompression. The vertical dashed line indicates the transition pressure.

**Figure 1**

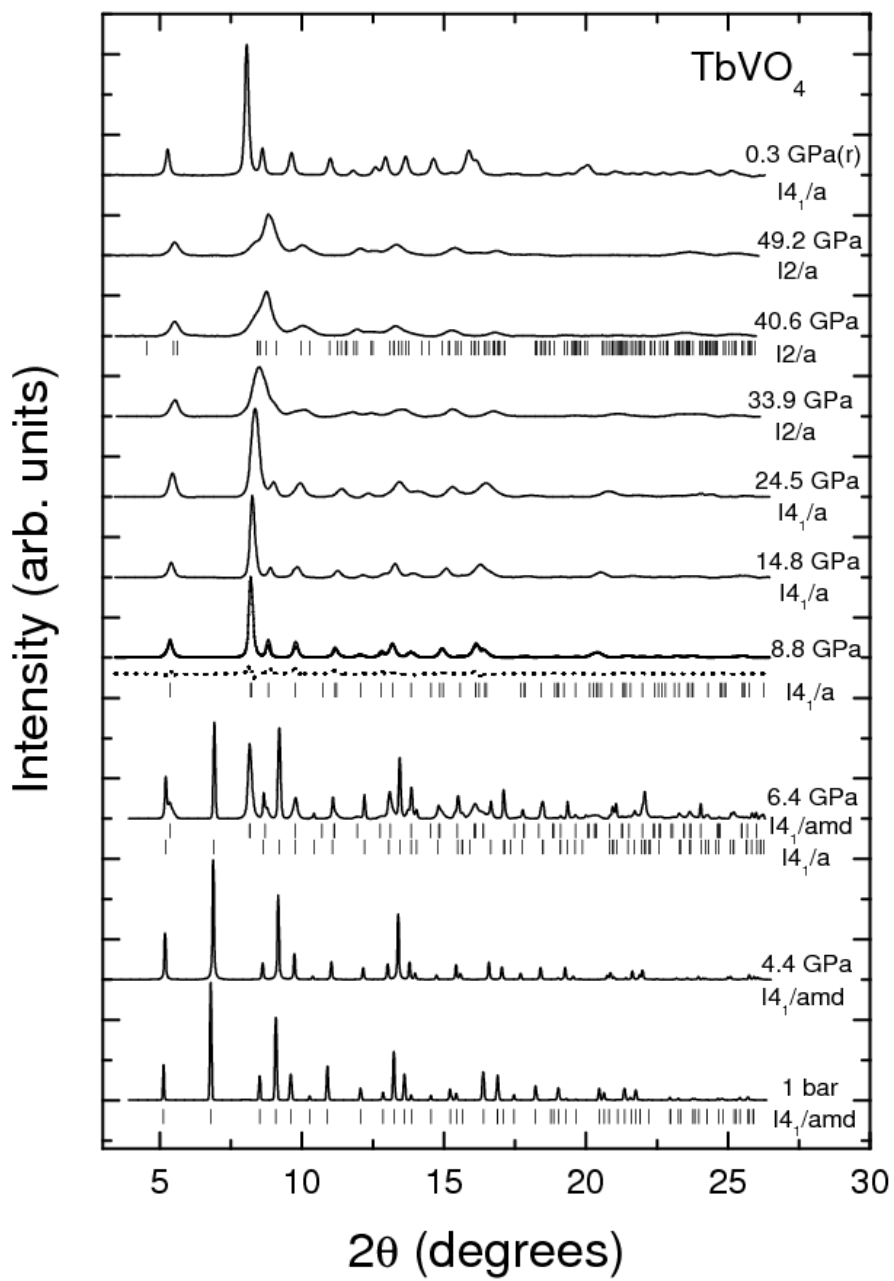

**Figure 2**

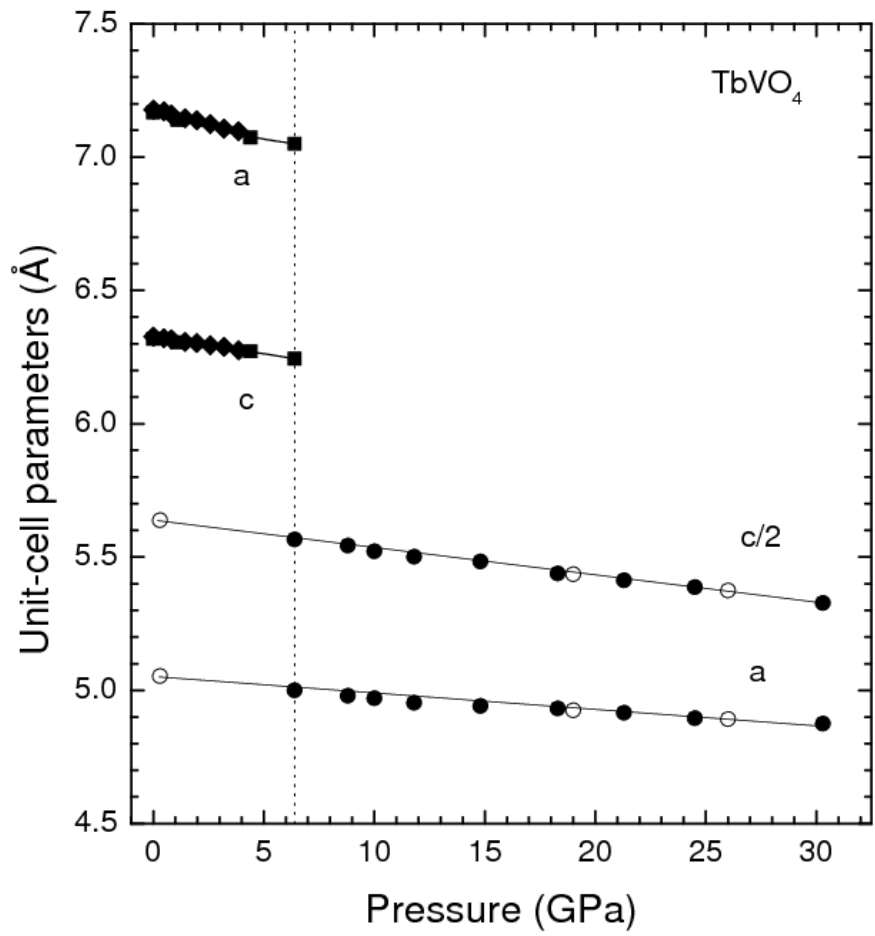

**Figure 3**

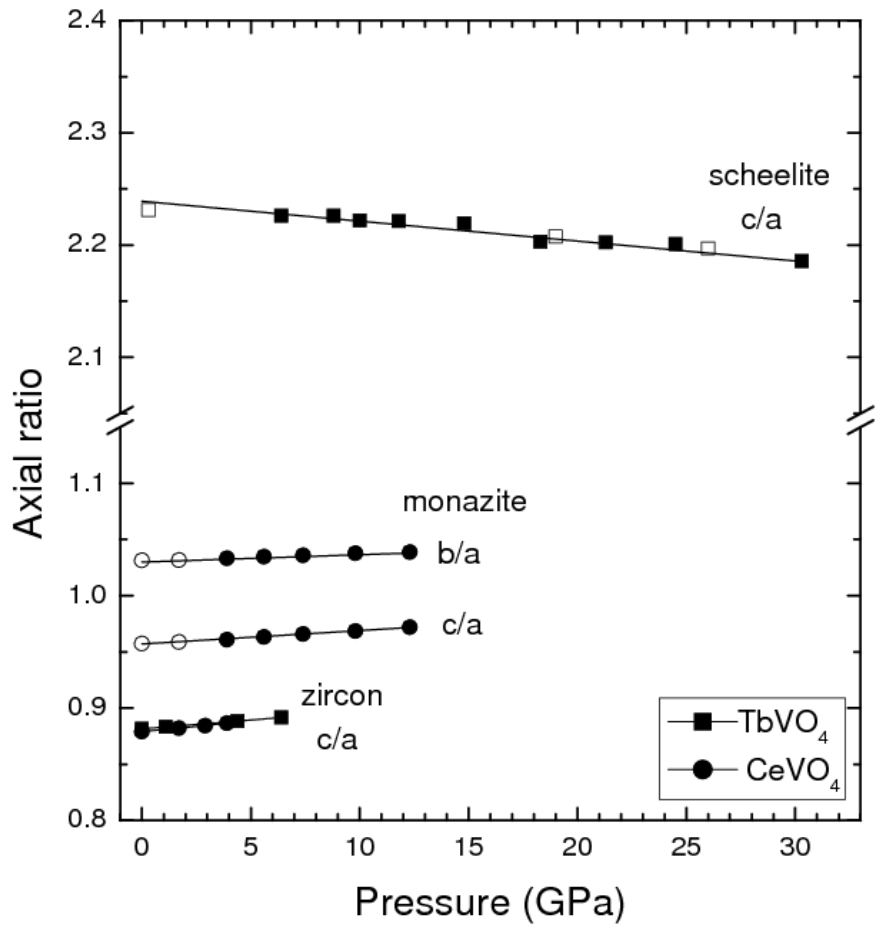

**Figure 4**

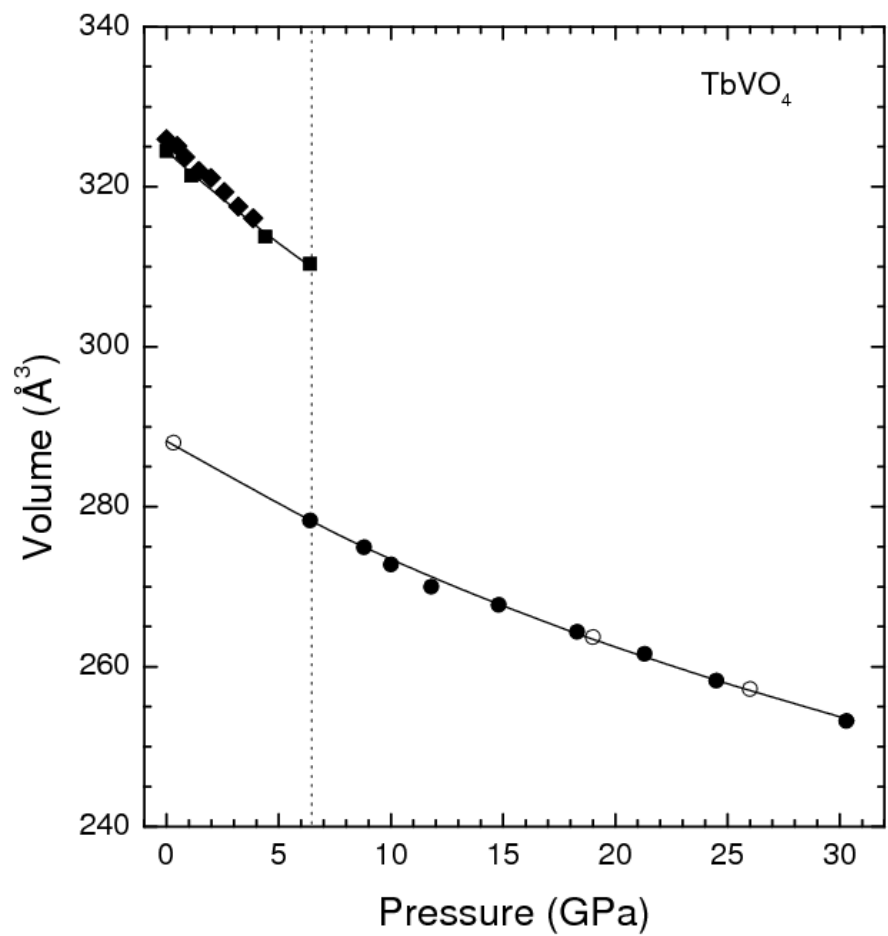

**Figure 5**

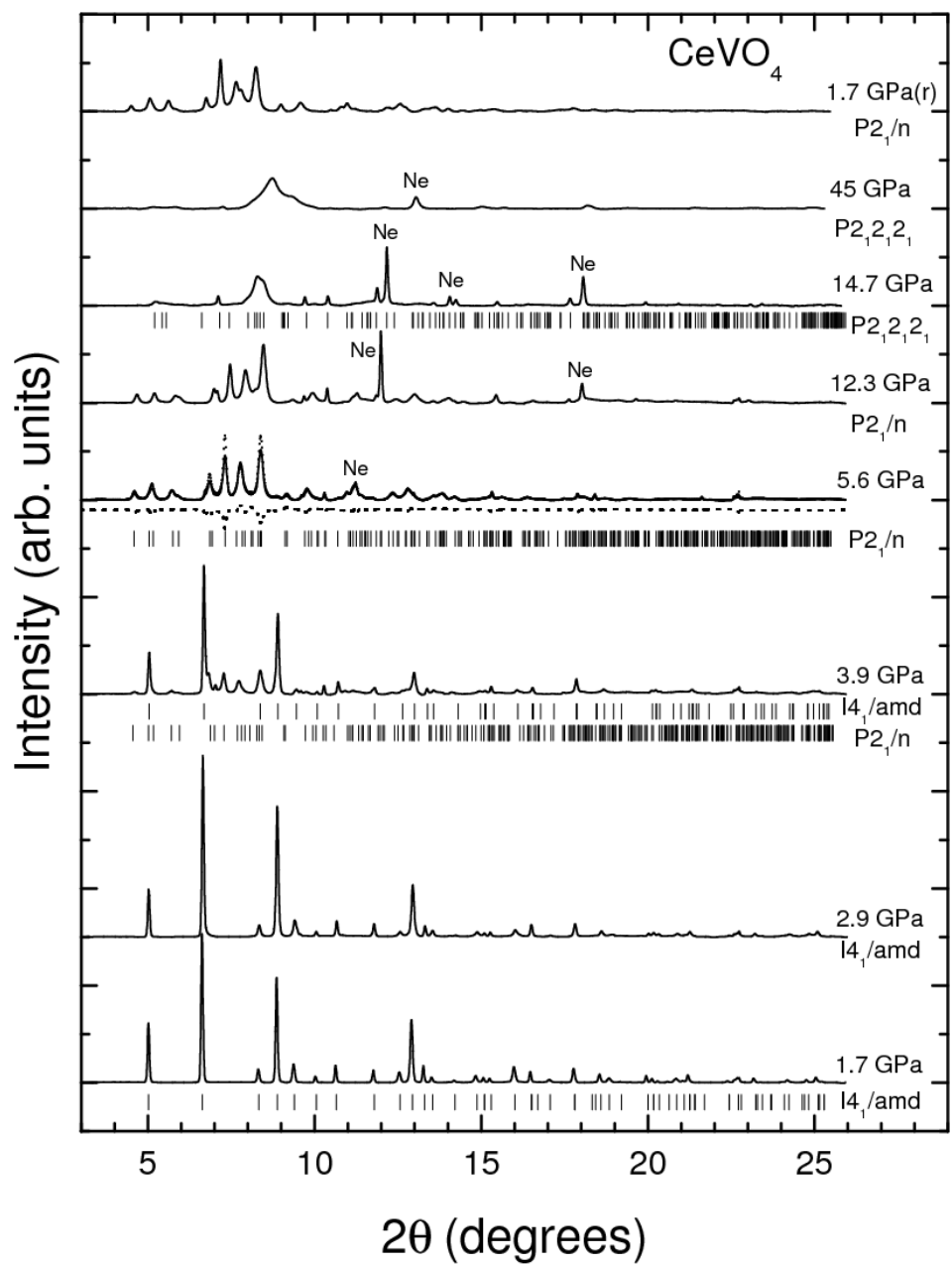

**Figure 6**

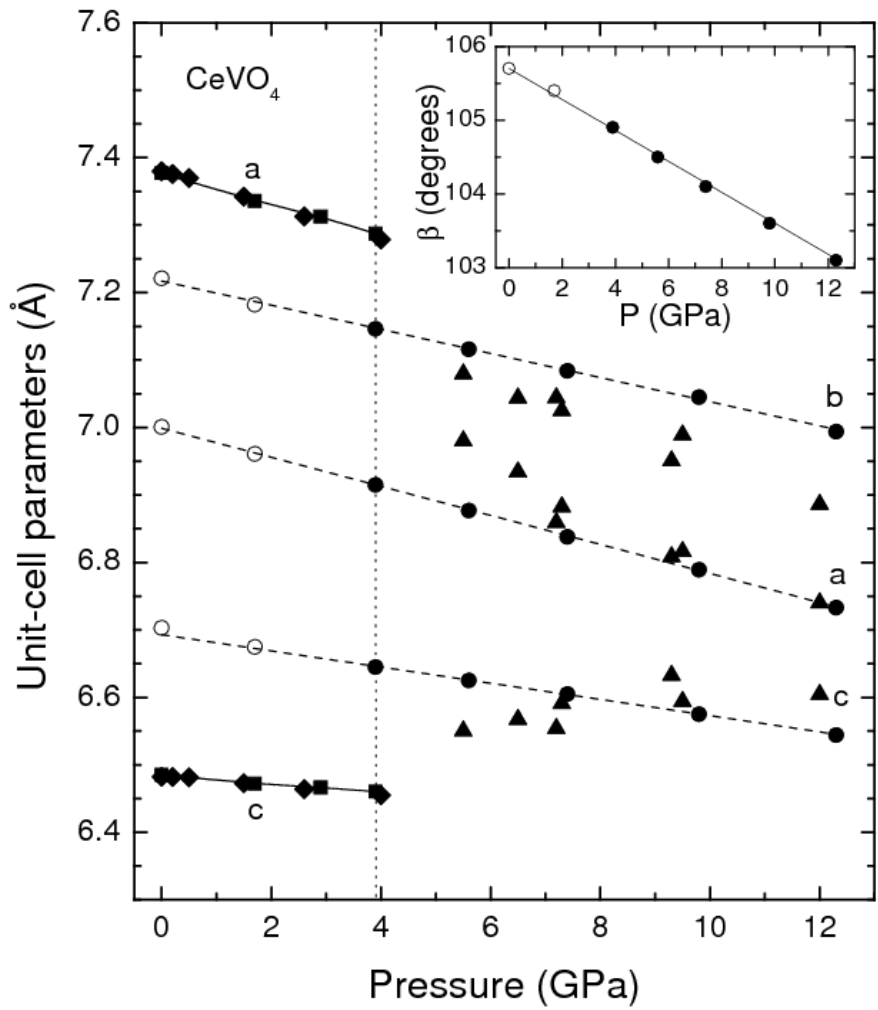

**Figure 7**

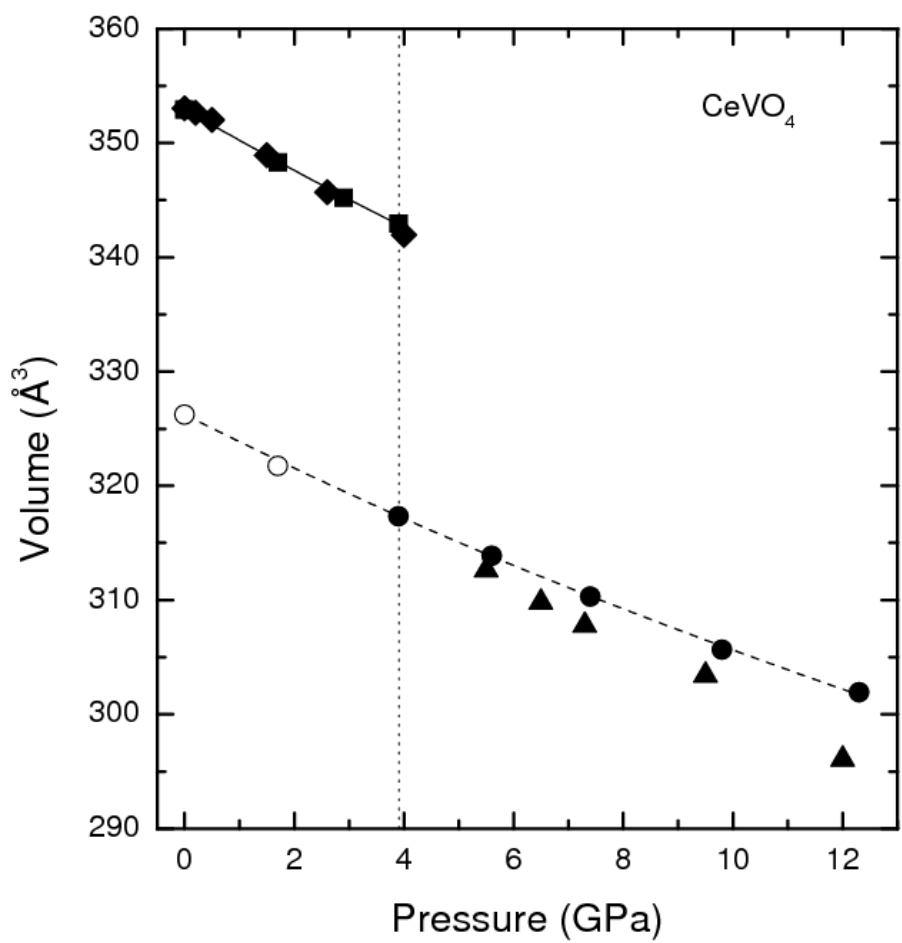